\documentclass[twocolumn,prl,amssymb,amsmath,preprintnumbers,aps,nofootinbib,superscriptaddress]{revtex4-1}
\usepackage{lineno}
\usepackage{bm,color,xcolor}
\usepackage{slashed} 
\usepackage{graphicx}
\usepackage{soul} 
\usepackage{amsmath}
\usepackage{multirow,tabularx}
\usepackage{tablefootnote}
\usepackage{mathtools}
\usepackage{appendix}
\usepackage{cancel}
\usepackage[export]{adjustbox}
\usepackage[colorlinks=true,urlcolor=blue,anchorcolor=blue,citecolor=blue,filecolor=blue,linkcolor=blue,menucolor=blue,linktocpage=true,pdfproducer=medialab,pdfa=true]{hyperref}

\begin{document}
\title{\boldmath \bf \Large 
High-Quality Axion Models with  the Anomalous $U(1)_X$ Gauge Symmetry 
}

\author{Hongkun Gao}
\affiliation{School of Sciences, Xi’an Technological University, Xi’an 710021, P. R. China}

\author{Tianjun Li}
\affiliation{School of Physics, Henan Normal University, Xinxiang 453007, P. R. China}
\affiliation{CAS Key Laboratory of Theoretical Physics, Institute of Theoretical Physics, Chinese Academy of Sciences, Beijing 100190, China}
\affiliation{School of Physical Sciences, University of Chinese Academy of Sciences, No. 19A Yuquan Road, Beijing 100049, China}

\author{Lina Wu}
\affiliation{School of Sciences, Xi’an Technological University, Xi’an 710021, P. R. China}

\author{Wenxing Zhang}
\affiliation{Department of Physics, Hebei University, Baoding 071002, China}

\begin{abstract}

We propose the generic high-quality axion models with anomalous $U(1)_X$ gauge symmetry and vector-like particles.
We briefly review the gauge anomaly cancellations via the Green-Schwarz mechanism, study the breaking of the $U(1)_X$ gauge symmetry, as well as derive the Nambu-Goldstone boson, Peccei-Quinn (PQ) axion, and axion decay constant in  general. The high-dimensional operators, which break the $U(1)_{PQ}$ global symmetry, have dimension eleven or higher due to the anomalous $U(1)_X$ gauge symmetry, and thus the axion quality problem is solved. 
In particular, unlike the high-quality axion models with anomaly free $U(1)$ gauge symmetry, we only need to introduce two pairs of vector-like particles. To be concrete, we present three specific models with two pairs of vector-like particles. We show that gauge anomalies in all three models can be canceled via the Green-Schwarz mechanism. To achieve gauge coupling unification, we need to introduce additional vector-like particles only in Model I.
We find that gauge coupling unification is achieved at the unification scale around $10^{16}$ GeV with a relative error of less than 1\%. Notably, gauge coupling unification in Model II is achieved naturally with the smallest relative error of 0.1\%.

\end{abstract}
\preprint{
}

\maketitle

\section{Introduction}

As we know, there exists a topological term for the Quantum Chromodynamics (QCD) theory in the Standard Model (SM) as follows
\begin{equation}
	{\cal L} \supset \frac{g^2_3}{16\pi^2} \,\overline{\theta}\, {\rm Tr}  (G_{\mu \nu} \tilde{G}^{\mu \nu})~,~
\end{equation}
where $g_3$ is the strong gauge coupling,  $G_{\mu \nu}$ is the gluon field strength, 
$\tilde{G}^{a,\mu \nu} = \frac{1}{2}\epsilon^{\mu \nu \alpha \beta} G_{\alpha \beta}^a$ is its dual. In particular, 
$\overline{\theta}$ is a dimensionless parameter, and it is the sum of the original $\theta$ and the contributions from the quark mass matrices.
This topological term violates the CP symmetry, and can generate the neutron electric dipole moment at the order of  $d_n \sim (10^{-16}\,\overline{\theta})\, e$-cm~\cite{Crewther:1979pi}.
The current experimental bound on $d_n$ is $|d_n| \leq 1.8 \times 10^{-26}\,e$-cm \cite{Abel:2020pzs}, and thus gives a stringent constraint on $|\overline{\theta}| \leq 10^{-10}$.
This fine-tuning problem is called the strong CP problem, which might indicate the new physics beyond the SM.

There are a few solutions to the strong CP problem, and the natural solution is the Peccei-Quinn (PQ) mechanism~\cite{Peccei:1977hh}. 
In this mechanism, we introduce a global $U(1)_{PQ}$ symmetry, which is anomalous under $SU(3)_C$ and can be broken by the instanton effect. Because the peudo Nambu-Goldstone boson (pNGB) from the $U(1)_{PQ}$ symmetry breaking cancels the strong CP phase $\overline{\theta}$ dynamically, the strong CP problem is solved elegantly~\cite{Weinberg:1977ma,Wilczek:1977pj}.

The original Weinberg--Wilczek axion~\cite{Weinberg:1977ma,Wilczek:1977pj} has already been excluded by various experiments, in
particular by the non-observation of the rare decay $K \rightarrow \pi + a$, for recent review, see Ref.~\cite{DiLuzio:2020wdo}.  
To evade the experimental bounds, two viable ``invisible'' axion models has been proposed: (1)~The Kim--Shifman--Vainshtein--Zakharov
(KSVZ) axion model, which introduces a SM singlet and a pair of extra
vector-like quarks that carry $U(1)_{PQ}$ charges while the SM fermions and
Higgs fields are neutral under $U(1)_{PQ}$ symmetry~\cite{Kim:1979if,Shifman:1979if}; (2)~The
Dine--Fischler--Srednicki--Zhitnitskii (DFSZ) axion model, in which a
SM singlet and one pair of Higgs doublets are introduced,
and the SM fermions and Higgs fields
are charged under $U(1)_{PQ}$ symmetry~\cite{Dine:1981rt,Zhitnitsky:1980tq}. Interestingly, the QCD axion emerges as a compelling cold dark matter candidate, and it yields the correct relic density if its decay constant $f_a$ is about $10^{11}$ GeV.

However, it is well-known that all the global symmetries in the nature can be violated by the quantum gravity effects, for example, black holes and worm holes. 
Thus, the $U(1)_{PQ}$ symmetry can be broken by the non-renormalizable high-dimensional operators suppressed by the reduced Planck scale. 
The $\overline{\theta}$ parameter will be shifted away from zero significantly, and then the strong CP problem cannot be explained, 
unless the coefficients of these non-renormalizable operators, which violate the $U(1)_{\rm PQ}$ symmetry, are extremely small, or the dimensions of 
these non-renormalizable operators are large. To be concrete, in order to preserve the solution to the strong CP problem,
one obtains that the coefficients of the dimension-five operators should be smaller than $ 10^{-50}$~\cite{Babu:2024udi, Babu:2024qzb}, or
the dimensions of the non-renormalizable operators are larger than or equal to 11 for  $f_a \ge 10^{11}$~\cite{Gherghetta:2025fip}. 
So there exists the axion quality problem due to the severe fine-tuning~\cite{Kamionkowski:1992mf, Holman:1992us, Barr:1992qq, Ghigna:1992iv}.

Various mechanisms, which can solve the axion quality problem, have been proposed: 
the accidental $U(1)_{\rm PQ}$ symmetries arising from the gauged $U(1)$ symmetries~\cite{Barr:1992qq, Fukuda:2017ylt, Qiu:2023los, Babu:2024udi, Babu:2024qzb} 
or non-Abelian gauge symmetries~\cite{DiLuzio:2020qio, Ardu:2020qmo}, 
the composite axions~\cite{Randall:1992ut, Lillard:2018fdt, Vecchi:2021shj, Lee:2018yak, Cox:2023dou, Cox:2021lii, Nakai:2021nyf, Gherghetta:2025fip}, 
the discrete gauged PQ symmetries from anomalous $U(1)_A$ gauge theory in string theory~\cite{Babu:2002ic, Barger:2004sf}, 
the cancellations of $\theta$ terms due to mirror symmetry~\cite{Berezhiani:2000gh, Hook:2019qoh}, 
the non-linearly realized discrete symmetries due to the $N$ copies of the SM~\cite{Hook:2018jle, Banerjee:2022wzk},  
the QCD axion from the five-dimensional parity odd gauge field~\cite{Choi:2003wr}, 
as well as the string axions~\cite{Svrcek:2006yi}, etc.

For the solutions to the axion quality problem via the accidental $U(1)_{\rm PQ}$ symmetries arising 
from the gauged $U(1)$ symmetries~\cite{Barr:1992qq, Fukuda:2017ylt, Qiu:2023los, Babu:2024udi, Babu:2024qzb}, 
we need to introduce eleven pairs of vector-like particles for the renormalizable theory, or equivalently introduce  
eleven pairs of vector-like particles for the non-renormalizable theory which arises from the renormaliable theory, 
where we count both the vector-like particles in the high-dimensional operators as well as 
the vector-like particles which are integrated out. To be concrete, to have dimension $n$ operators
$XF \overline{XF} \Phi^{n-3}/M_*^{n-4}$ with $n\ge 4$ from the renormalizable theory, we need to introduce $n-3$ pair of vector-like particles.
It is trivial for $n=4$, and has been proved for $n=8$ explicitly in Ref.~\cite{Babu:2024qzb}.

In this paper, to construct the high quality axion model with less number of the vector-like particles, we consider the anomalous $U(1)_X$ gauge symmetry whose gauge anomalies are canceled by the Green-Schwarz mechanism~\cite{Green:1984sg}. 
First, we propose the generic high-quality axion models with anomalous $U(1)_X$ gauge symmetry and vector-like particles.
We briefly review the gauge anomaly cancellations via the Green-Schwarz mechanism, study the breaking of the $U(1)_X$ gauge symmetry, as well as derive the Nambu-Goldstone boson, Peccei-Quinn (PQ) axion, and axion decay constant in  general. The high-dimensional operators, which break the $U(1)_{PQ}$ global symmetry, have dimension eleven or higher due to the anomalous $U(1)_X$ gauge symmetry, and thus the axion quality problem is solved. 
Second, we propose three models. In Model I, we introduce two pairs of the vector-like particles which belong to the fundamental and anti-fundamental representations of $SU(5)$ model.  We cancel the gauge anomalies via the Green-Schwarz mechanism.  To achieve gauge coupling unification, we introduce the TeV-scale vector-like particles. Also, we discuss the variation of this model. In Model II, we introduce two pair of vector-like particles which belong to the anti-symmetric representation and its Hermitian conjugate of the flipped $SU(5)$ model. The gauge anomalies can be canceled by Green-Schwarz mechanism. In particular, the gauge coupling unification can be achieved naturally.
By the way, the SM singlets in the anti-symmetric representation and
its Hermitian conjugate can be considered as the right-handed neutrinos, or we do not need to introduce them.
In Model III, we introduce two pairs of vector-like particles: 
the quantum numbers for one pair of vector-like particles are the same as the 
quark doublet and its Hermitian conjugate, and  the quantum numbers for the other pair of vector-like particles are the same as the righ-handed down-type quark and its Hermitian conjugate. 
Also, the gauge anomalies can be canceled by Green-Schwarz mechanism.
Moreover, these vector-like particles can obtain the masses from the renormalizable Yukawa couplings. Interestingly,  if these vector-like particles obtain the masses from the non-renormalizable dimension-five Yukawa couplings, their masses can be around TeV scale. We show that we can achieve the gauge coupling unification as well.

\section{The Generic High Quality Axion Models with Anomalous $U(1)_X$ Gauge Symmetry}

First,  let us introduce our convention, which is similar to the supersymmetric SM, {i.e.}, 
all the chiral fermions in the left-handed format. We introduce
the vector-like particles whose quantum numbers are the same as those
of the SM fermions and their Hermitian conjugates.
Their quantum numbers under 
$SU(3)_C \times SU(2)_L \times U(1)_Y$ and their
contributions to one-loop beta functions, $\Delta b \equiv (\Delta
b_1, \Delta b_2, \Delta b_3)$ are~\cite{Barger:2007qb}
\begin{eqnarray}
	&& XQ + {\overline{XQ}} = {\mathbf{(3, 2, {1\over 6}) + ({\bar 3}, 2,
			-{1\over 6})}}\,, \quad \Delta b =({2\over 15}, 2, {4\over 3})\,; \nonumber \\ 
	&& XU + {\overline{XU}} = {\mathbf{ ({3},
			1, {2\over 3}) + ({\bar 3},  1, -{2\over 3})}}\,, \quad \Delta b =
	({16\over 15}, 0, {2\over 3})\,;\nonumber \\ 
	&& XD + {\overline{XD}} = {\mathbf{ ({3},
			1, -{1\over 3}) + ({\bar 3},  1, {1\over 3})}}\,, \quad \Delta b =
	({4\over 15}, 0, {2\over 3})\,;\nonumber \\  
	&& XL + {\overline{XL}} = {\mathbf{(1,  2, {1\over 2}) + ({1},  2,
			-{1\over 2})}}\,, \quad \Delta b = ({6\over 15}, {2\over 3}, 0)\,;\nonumber \\ 
	&& XE + {\overline{XE}} = {\mathbf{({1},  1, {1}) + ({1},  1,
			-{1})}}\,, \quad \Delta b = ({12\over 15}, 0, 0)\,, \nonumber\\
	&& XN + {\overline{XN}} = {\mathbf{({1},  1, {0}) + ({1},  1,
			{0})}}\,, \quad \Delta b = (0, 0, 0)\,. \nonumber	\\
	&& XG  = {\mathbf{({8},  1, {0})}\,, \quad \Delta b = (0, 0, 2)}\,; \nonumber \\	
	&& XW  = {\mathbf{({1},  3, {0})}\,, \quad \Delta b = (0, {4\over 3}, 0)}\,; \nonumber \\	
	&& XB  = {\mathbf{({1},  1, {0})}\,, \quad \Delta b = (0, 0, 0)}\,. \nonumber
\end{eqnarray}

We define 
\begin{eqnarray}
	&& XF\equiv XD+{\overline{XL}}~,~{\overline{XF}} \equiv \overline{XD}+XL\,; \nonumber \\ 
	&& XT \equiv XQ+\overline{XD}+\overline{XN}~,~{\overline{XT}} \equiv {\overline{XQ}}+XD+XN\,. \nonumber
\end{eqnarray}
Thus, $XF$ and ${\overline{XF}}$ belong to the fundamental and anti-fundamental representations of $SU(5)$ model, and $XT$ and ${\overline{XT}}$ belong to the
anti-symmetric representation and its Hermitian conjugate of flipped $SU(5)$ model.
Their quantum numbers under 
$SU(5)$/flipped $SU(5)$ models, and their contributions to one-loop beta functions, $\Delta b \equiv (\Delta
b_1, \Delta b_2, \Delta b_3)$ are
\begin{eqnarray}
	&& XF+ {\overline{XF}} = {\mathbf{5+ {\bar 5}}}\,, \quad \Delta b =( {2\over 3}, {2\over 3}, {2\over 3})\,; \nonumber \\ 
	&& XT+ {\overline{XT}} = {\mathbf{(10, 1)+ (\overline{10}}}, -1)\,, \quad \Delta b =( {2\over 5}, 2, 2)\,. \nonumber
\end{eqnarray}

In general, we introduce two pairs of vector-like particles $(Xf, \overline{Xf})$ and
$(Xf', \overline{Xf'})$. To realize the high quality axions and cancel the gauge anomalies
via Green-Schwarz mechanism, $(Xf, \overline{Xf})$ and
	$(Xf', \overline{Xf'})$ cannot be singlets under $SU(3)_C\times SU(2)_L$ and are neutral  under $U(1)_Y$. 
To break the $U(1)_X$ gauge symmetry, we introduce two Higgs fields $\Phi_1$ and $\Phi_2$.  
We assume that the $U(1)_X$ charges for $Xf/\overline{Xf}$, $Xf'/\overline{Xf'}$, $\Phi_1$ and $\Phi_2$ are $p$, $-q$, $q_1$, 
and $q_2$. For renormalizable theory, we have $q_1=\pm 2p$, and $q_2=\pm 2q$. While for the non-renormalizable theory 
with dimension-five operators, we have $q_1=\pm p$, and $q_2=\pm q$. 
To be concrete, we further assume that $p$ and $q$ are positive integers 
and coprime, and  $q_1=- 2p$, and $q_2= 2q$.
The $U(1)_X \times U(1)_{PQ}$ quantum numbers for vector-like particles and Higgs fields are given in Table~\ref{QN}. The discussions for the other 
values of $p$, $-q$, $q_1$, and $q_2$ are similar.

\begin{table}[t] 
	\label{tab:example}
	\small
	\centering
	\begin{tabular}{|c|c|c|}
		\hline
		\textbf{Particles} & \textbf{\boldmath{$U(1)_X$} } & ~\textbf{\boldmath{$U(1)_{PQ}$}}~  \\ 
		\hline
		$(Xf,\overline{Xf})$ & $p$ & $q$ \\
		\hline
		$(Xf',\overline{Xf'})$ & $-q$ & $p$ \\
		\hline
		$\Phi_1$& $q_1=-2p$ & $-2q$ \\
		\hline
		$\Phi_2$ & $q_2=2q$ & $-2p$ \\
		\hline\hline 
	\end{tabular}
	\caption{The $U(1)_X \times U(1)_{PQ}$ quantum numbers for vector-like particles and Higgs fields. In particular, $p$ and $q$ are coprime positive integers.}
	\label{QN}
\end{table}

It is easy to show that the $\left[U(1)_Y  U(1)^2_X \right]$ gauge anomaly is canceled, and
we have the gauge anomalies $A_1\equiv \left[U(1)_Y^2 U(1)_X \right]$, 
$A_2\equiv \left[SU(2)^2_L U(1)_X \right]$, $A_3 \equiv \left[SU(3)^2_C U(1)_X \right]$,
$A_{1X} \equiv \left[U(1)_X^3\right]$, and gravity anomaly  $A_{\rm gravity} \equiv \left[{\rm Gravity}^2 U(1)_X \right]$. These anonalies can be canceled via 
the Green-Schwarz mechanism.

In weakly coupled heterotic string model building, in general we have  
an anomalous $U(1)_X$ gauge symmetry with its anomalies canceled by the
Green--Schwarz mechanism~\cite{Green:1984sg}.  For Type II orientifold string model
building, we have more than one anomalous $U(1)_X$ gauge symmetries
whose anomalies can be cancelled by the generalized Green--Schwarz
mechanism~\cite{Aldazabal:2000dg}. The Green--Schwarz anomaly
cancellation conditions from an effective theory point of view 
are~\cite{Banks:1991xj, Ibanez:1991hv, Ibanez:1991pr}
\begin{eqnarray}
	\frac{A_{i}}{k_{i}}= \frac{A_{1X}}{k_{1X}}
	=\frac{A_{\rm gravity}}{12}=\delta_{GS}~,~\,
\end{eqnarray}
where  $i=1,~2,~3$. Also,
$k_i$ and $k_{1X}$ are the levels of the corresponding Kac--Moody algebra, and $\delta_{GS}$
is a constant which is not specified by low-energy theory alone.  For a
non-Abelian group, $k_i$ is a positive integer, while for the $U(1)$ gauge
symmetry, $k_i$ needs not be an integer.  To be concrete, $k_1$ and $k_{1X}$
are positive real numbers, and $k_2$ and $k_3$ are positive integers. 
Thus, the anomaly cancellation conditions for $A_1$ and $A_{1X}$ 
are not very useful from an effective low energy theory point of view. 
Also, the anomalies $A_{1X}$ and $A_{\rm gravity}$ have an arbitrariness due to 
the normalization of $U(1)_X$, and they can be easily canceled by introducing
SM singlet fermions which are charged under $U(1)_X$. Thus, we shall not consider them here.
From the four-dimensional  effective field theory point of view, we can cancel
the gauge anomalies  $A_1\equiv \left[U(1)_Y^2 U(1)_X \right]$, 
$A_2\equiv \left[SU(2)^2_L U(1)_X \right]$, $A_3 \equiv \left[SU(3)^2_C U(1)_X \right]$,
$A_{1X} \equiv \left[U(1)_X^3\right]$, and gravity anomaly  $A_{\rm gravity} \equiv \left[{\rm Gravity}^2 U(1)_X \right]$ by introducing two-form fields and its dual singlet scalars.
By the way, all the other gauge anomalies 
such as $G_i G_j G_k$ and $[U(1)_X]^2 \times G_i$ must vanish, and we can easily confirm 
that these conditions are satisfied.

The Lagrangian for the Yukawa Couplings is
\begin{eqnarray}
	-{\cal L} = y_{Xf} Xf \overline{Xf} \Phi_1 + y_{Xf'} Xf' \overline{Xf'} \Phi_2~.~\label{eq:Ycoupling}
\end{eqnarray}

And the Higgs potential at the renormalizable level is
\begin{eqnarray}
V&=& M_1^2 |\Phi_1|^2 + M_2^2 |\Phi_2|^2 + \frac{\lambda_1}{4} |\Phi_1|^4
+ \frac{\lambda_2}{4} |\Phi_2|^4 \nonumber \\
&& + \lambda_{12} |\Phi_1|^2 |\Phi_2|^2 ~.~\,
\label{Higgs-Potential}
\end{eqnarray}

After $\Phi_i$ obtain the Vacuum Expectation Values (VEVs), we parametrize $\Phi_i$ as
\begin{eqnarray}
\Phi_i=\frac{1}{\sqrt 2} \left(\phi_i + f_i\right) e^{ib_i/f_i} ~,~\,
\end{eqnarray}
where $\phi_i$ are the radial components, $b_i$ are the axial components, and  
$f_i$ are the VEVs of $\Phi_i$. The domains of $b_i/f_i$ 
are $[0, 2\pi)$. 

For the axial components, the $U(1)_X$ gauge symmetry becomes shift symmetry, and the transformations of axial components are
\begin{eqnarray}
b_i \rightarrow b_i + q_i f_i \epsilon~.~\, 
\end{eqnarray}	

For the Higgs potential in Eq.~(\ref{Higgs-Potential}), we have two global symmetries from the phases of $\Phi_i$:
one is the $U(1)_X$ gauge symmetry, and the other is the $U(1)_{PQ}$ symmetry. 
Let us study it in details. 
From the kinetic terms of $\Phi_i$, we obtain
\begin{eqnarray}
{\cal L} &=& |D_{\mu} \Phi_1|^2 + |D_{\mu} \Phi_2|^2 
\nonumber \\&&
\supset \frac{1}{2} \partial^{\mu} b_1 \partial_{\mu} b_1 + \frac{1}{2} \partial^{\mu} b_2 \partial_{\mu} b_2 
\nonumber \\&&
-\left(  q_1 f_1 \partial^{\mu} b_1 + q_2 f_2 \partial^{\mu} b_2\right) 
g_X A_{X\mu} 
\nonumber \\&&
+ \frac{1}{2} g_X^2\left(q_1^2 f_1^2 +q_2^2 f_2^2\right) A_X^{\mu} A_{X\mu} ~,~\, 
\label{Kinetic-Terms} 
\end{eqnarray}
where $A_{X\mu}$ is gauge field and $g_X$ is gauge coupling of the $U(1)_X$ gauge theory.
The mass of the $U(1)_X$ gauge boson is
\begin{eqnarray}
M_X=g_X^2\left(q_1^2 f_1^2 +q_2^2 f_2^2\right)~.~\,
\end{eqnarray}

We redefine the axial components as
\begin{eqnarray}
	\left(
	\begin{array}{cc}
		a   \\
		G
	\end{array}
	\label{eq:decomp}
	\right)=
	\frac{1}{\sqrt{q_1^2 f_1^2 + q_2^2f_2^2 }}\left(
	\begin{array}{cc}
		q_2 f_2   &  -q_1 f_1   \\
		q_1 f_1  &   q_2 f_2
	\end{array}
	\right)
	\left(
	\begin{array}{cc}
		 b_1   \\
		b_2
	\end{array}
	\right)\ .
\end{eqnarray}
From Eq.(\ref{Kinetic-Terms}), we obtain that the axial field $ G$ is 
the  Nambu-Goldstone boson for the $U(1)_X$ gauge symmetry breaking, 
while the gauge invariant axial field $a$ is the PQ axion.

Due to the QCD anomaly, we obtain the anomalous couplings for $b_i$
\begin{eqnarray}
	{\cal L} = \frac{g_3^2}{32\pi^2} 
	\left(\frac{b_1}{ f_1} + \frac{b_2}{ f_2}\right)
	G\tilde{G} \ .
\end{eqnarray}
And then we obtain the axion decay constant 
\begin{eqnarray}
f_a = \frac{f_1 f_2 \sqrt{p^2 f_1^2 + q^2 f_2^2}}{pf_1^2+qf_2^2}~.~\,
\end{eqnarray}
Because $U(1)_X$ is anomalous gauge symmetry, the decay constants in our models are different from those in Refs.~\cite{Fukuda:2017ylt, Qiu:2023los, Babu:2024udi, Babu:2024qzb}. Of course, if we consider the anomaly free $U(1)_X$ gauge symmetry,
for example, we introduce  $q$ pairs of vector-like particles $(Xf, \overline{Xf})$,
and $p$ pairs of the vector-like particles $(Xf', \overline{Xf'})$, we obtain
the same axion decay constant as in Refs.~\cite{Fukuda:2017ylt, Qiu:2023los, Babu:2024udi, Babu:2024qzb}
\begin{eqnarray}
	f_a = \frac{f_1 f_2 }{\sqrt{p^2 f_1^2 + q^2 f_2^2}}~.~\,
\end{eqnarray}

It is well-known that quantum gravitational effects, which are associated with black holes, worm holes, etc.,
are believed to violate all the global symmetries, while they respect all the
gauge symmetries~\cite{Hawking:1987mz, Lavrelashvili:1987jg, Giddings:1987cg}. 
The non-renormalizable operators with the lowest dimension for Higgs fields $\Phi_1$ and $\Phi_2$, 
which preserve the $U(1)_X$ gauge symmetry whike violate the $U(1)_{PQ}$ symmetry, are 
\begin{eqnarray}
	{\cal L} \supset  \frac{1}{p!q!} \frac{\Phi_1^q \Phi_2^p}{M_{\rm Pl}^{p+q-4}} + {\rm H. C.} ~.~\,
\end{eqnarray}
Thus, we obtain $p+q\ge 11$. For example, we can have 
\begin{eqnarray}
(p, q) &=&(1, 10), ~(2, 9),~(3, 8),~(4, 7),~(5, 6),~(6, 5),
\nonumber \\&& ~(7, 4),~(8, 3),~(9, 2),~(10, 1)
~.~\,
\end{eqnarray}

\section{Model I with Vector-Like Particles $(XF, \overline{XF})$ and $(XF', \overline{XF'})$ }

In Model I, we introduce two paris of vector-like particles
$(XF, \overline{XF})$ and $(XF', \overline{XF'})$, as well as 
the Higgs fields $\Phi_1$ and $\Phi_2$.
The $U(1)_X \times U(1)_{PQ}$ quantum numbers for vector-like particles and Higgs fields are given in Table~\ref{QN-M1}.

\begin{table}[t] 
	\small
	\centering
	\begin{tabular}{|c|c|c|}
		\hline
		\textbf{Particles} & \textbf{\boldmath{$U(1)_X$} } & ~\textbf{\boldmath{$U(1)_{PQ}$}}~  \\ 
		\hline
		$(XF,\overline{XF})$ & $p$ & $q$ \\
		\hline
		$(XF',\overline{XF'})$ & $-q$ & $p$ \\
		\hline
		$\Phi_1$& $-2p$ & $-2q$ \\
		\hline
		$\Phi_2$ & $2q$ & $-2p$ \\
		\hline\hline 
	\end{tabular}
	\caption{The $U(1)_X \times U(1)_{PQ}$ quantum numbers for vector-like particles and Higgs fields in Model I. 
		In particular, $p$ and $q$ are coprime positive integers.}
	\label{QN-M1}
\end{table}

First, we discuss the cancellation conditions for the gauge anomalies via
the Green-Schwarz mechanism. The gauge anomalies in Model I are
\begin{eqnarray}
    A_1=A_2=A_3=p-q~.~\,
\end{eqnarray}
Thus, we obtain 
\begin{eqnarray}
	k_1=k_2=k_3=p-q~.~\,
\end{eqnarray}                      

\begin{figure}[ht]
    \centering 
    \includegraphics[width=0.9\linewidth]{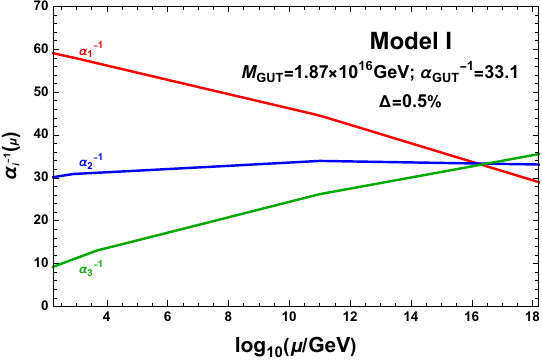}
    \caption{The evolution of two-loop gauge couplings in the Model I with vector-like particles $(XF, \overline{XF})$ and $(XF', \overline{XF'})$. The masses of the vector-like particles are set as follows: $M_{XW}=M_{(XL,\overline{XL})}=800$ GeV for $XW$ and $(XL, \overline{XL})$, $M_{XB}=M_{XG}=5$ TeV for $XB$ and $XG$, and $M_{(XF,\overline{XF})}=M_{(XF',\overline{XF'})}=10^{11}$ GeV for $(XF,\overline{XF})$ and $(XF',\overline{XF'})$.}
    \label{fig:RGE_MI}
\end{figure}

Second, to achieve the gauge coupling unification, we introduce $XB$, $XW$, $XG$, and $(XL, \overline{XL})$ 
around the TeV scale. Because the vector-like particles $(XF, \overline{XF})$ and $(XF', \overline{XF'})$ form the complete $SU(5)$ multiplets, 
they do not affect the gauge coupling unification. For simplicity, we assume their
masses around $10^{11}$ GeV via the renormalizable Yukawa couplings.  Following the convention in Refs.~\cite{Chen:2017rpn,He:2021kbj,Li:2022cqk,Mansha:2022pnd,Mansha:2023kwq},  we numerically solved the two-loop renormalization group equations (RGEs) for the gauge couplings by setting the central value of top quark pole mass as $m_t = 173.34\pm 0.27(\rm stat)\pm 0.71(\rm syst) $ GeV \cite{1403.4427} and taking the strong coupling constant at reference $Z$ boson mass as $\alpha_3(M_Z^2)= 0.1179\pm 0.0009 $ \cite{dEnterria:2022hzv}. The evolution of the two-loop gauge couplings is presented in Fig.~\ref{fig:RGE_MI}.  
The unification condition is defined as $\alpha^{-1}_\text{GUT}\equiv \alpha^{-1}_{1}=(\alpha^{-1}_{2}+\alpha^{-1}_{3})/2$ and the unification is achieved at the GUT scale $M_\text{GUT}=1.9\times 10^{16}$ GeV. The relative error $\Delta\equiv |\alpha^{-1}_{1}-\alpha^{-1}_{2}|/\alpha^{-1}_{1}$ quantifies the fractional deviation of $\alpha_{2,3}$ from $\alpha_1$.
Evidently, TeV-scale vector-like particles ($XB$, $XW$, $XG$, and $(XL, \overline{XL})$) must be incorporated to mitigate this relative error, and these particles exhibit a small mass splitting. Specifically, when $XW$ is introduced at 800 GeV with the remaining particles introduced at 5 TeV, the unification is achieved with a relative error of 0.9\%. However, if the mass of both $XW$ and $(XL, \overline{XL})$ are adjusted to $M_{XW}=M_{(XL,\overline{XL})}=800$ GeV,  while $ XB$ and $XG$ remain fixed at $M_{XB}=M_{XG}=5$ TeV, the relative error is further reduced to 0.5\%. For the remaining vector-like particles $(XF, \overline{XF})$ and $(XF', \overline{XF'})$, the impact of their masses on the relative error is negligible. Consequently, within the framework of this model, the masses of these particles are set to $M_{(XF,\overline{XF})}=M_{(XF',\overline{XF'})}=10^{11}$ GeV.

Third, if we do not consider gauge coupling unification, for simplicity, we can introduce
vector-like particles  $(XD', \overline{XD'})$ instead of  $(XF', \overline{XF'})$.
And then  the gauge anomalies  are
\begin{eqnarray}
	A_1=p-\frac{2}{5}q~,~~A_2=p~,~~A_3=p-q~.~\,
\end{eqnarray}
Thus, we obtain 
\begin{eqnarray}
	k_1=p-\frac{2}{5}q~,~~k_2=p~,~~k_3=p-q~.~\,
\end{eqnarray}

\section{Model II with Vector-Like Particles $(XT, \overline{XT})$ and $(XT', \overline{XT'})$}

In Model II, we introduce two paris of vector-like particles
$(XT, \overline{XT})$ and $(XT', \overline{XT'})$, as well as 
the Higgs fields $\Phi_1$ and $\Phi_2$.
The $U(1)_X \times U(1)_{PQ}$ quantum numbers for vector-like particles and Higgs fields are given in Table~\ref{QN-M2}.

\begin{table}[t] 
	\label{tab:example}
	\small
	\centering
	\begin{tabular}{|c|c|c|}
		\hline
		\textbf{Particles} & \textbf{\boldmath{$U(1)_X$} } & ~\textbf{\boldmath{$U(1)_{PQ}$}}~  \\ 
		\hline
		$(XT,\overline{XT})$ & $p$ & $q$ \\
		\hline
		$(XT',\overline{XT'})$ & $-q$ & $p$ \\
		\hline
		$\Phi_1$& $-2p$ & $-2q$ \\
		\hline
		$\Phi_2$ & $2q$ & $-2p$ \\
		\hline\hline 
	\end{tabular}
	\caption{The $U(1)_X \times U(1)_{PQ}$ quantum numbers for vector-like particles and Higgs fields in Model II. 
		In particular, $p$ and $q$ are coprime positive integers.}
	\label{QN-M2}
\end{table}

First, we discuss the cancellation conditions for the gauge anomalies via
the Green-Schwarz mechanism. The gauge anomalies in Model II are
\begin{eqnarray}
	5A_1=A_2=A_3=3(p-q)~.~\,
\end{eqnarray}
Thus, we obtain 
\begin{eqnarray}
	5k_1=k_2=k_3=3(p-q)~.~\,
\end{eqnarray}

\begin{figure}
    \centering
    \includegraphics[width=0.9\linewidth]{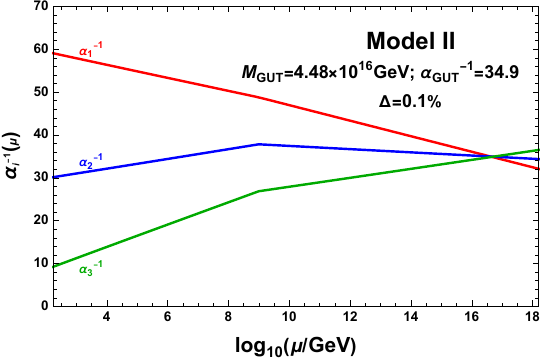}
    \caption{The evolution of two-loop gauge couplings in the Model II with vector-like particles $(XT, \overline{XT})$ and $(XT', \overline{XT'})$. The masses of these particles are set as $M_{(XT, \overline{XT})}=M_{(XT', \overline{XT'})}=10^9$ GeV. }
    \label{fig:RGE_MII}
\end{figure}

Second, we assume the Yukawa couplings $y_{xf}$ and $y_{xf}'$ in Eq.~\eqref{eq:Ycoupling} are of $\mathcal{O}(10^{-2})$,
and then
the vector-like particles $(XT, \overline{XT})$ and $(XT', \overline{XT'})$ have masses
around $ 10^{9}$ GeV.  
Interestingly, we can achieve gauge coupling unification
without introducing new particles. In Fig.~\ref{fig:RGE_MII}, after numerically solving the two-loop RGEs using input parameters consistent with earlier analyses, unification is achieved at $M_{\text{GUT}}=4.5\times 10^{16}$ GeV with a mere 0.1\% relative error. Notably, the relative error $\Delta$ of this model is substantially lower than the 0.5\%-0.9\% relative errors observed in the other two Models. When the vector-like particles $(XT, \overline{XT})$ and $(XT', \overline{XT'})$ are introduced at $10^9$ GeV, the RGE evolution curves of the gauge couplings exhibit a characteristic deflection. This deflection arises from the non-trivial beta function contributions of these particles. Unlike the TeV-scale particles $(XQ,\overline{XQ})$ and $(XD',\overline{XD'})$ in Model III, a key advantage of this model is that the minimal relative error can still be achieved without the need for mass splitting.   

\section{Model III with Vector-Like Particles $(XQ, \overline{XQ})$ and
	$(XD', \overline{XD'})$}

In Model III, we introduce two paris of vector-like particles
$(XQ, \overline{XQ})$ and $(XD', \overline{XD'})$,
as well as the Higgs fields $\Phi_1$ and $\Phi_2$.
The $U(1)_X \times U(1)_{PQ}$ quantum numbers for vector-like particles and Higgs fields are given in Table~\ref{QN-M3}. For the renormalizabe model,
we have $q_1=-2p$ and $q_2=-2q$. While for the non-renormalizable model,
we have  $q_1=-p$ and $q_2=-q$.

\begin{table}[t] 
	\label{tab:example}
	\small
	\centering
	\begin{tabular}{|c|c|c|}
		\hline
		\textbf{Particles} & \textbf{\boldmath{$U(1)_X$} } & ~\textbf{\boldmath{$U(1)_{PQ}$}}~  \\ 
		\hline
		$(XQ,\overline{XQ})$ & $p$ & $-q$ \\
		\hline
		$(XD',\overline{XD'})$ & $q$ & $p$ \\
		\hline
		$\Phi_1$ &  $q_1$ &  $-q_2$ \\
		\hline
		$\Phi_2$ &  $q_2$ &  $ q_1$ \\
		\hline\hline 
	\end{tabular}
	\caption{The $U(1)_X \times U(1)_{PQ}$ quantum numbers for vector-like particles and Higgs fields in Model III. 
		In particular, $p$ and $q$ are coprime positive integers.}
	\label{QN-M3}
\end{table}

First, we discuss the cancellation conditions for the gauge anomalies via
the Green-Schwarz mechanism. The gauge anomalies in Model III are
\begin{eqnarray}
	A_1=\frac{1}{5}p+\frac{2}{5}q~,~~A_2=3p~,~~A_3=2p+q~.~\,
\end{eqnarray}
Thus, we obtain 
\begin{eqnarray}
	k_1=\frac{1}{5}p+\frac{2}{5}q~,~~k_2=3p~,~~k_3=2p+q~.~\,
\end{eqnarray}

\begin{figure}
    \centering
    \includegraphics[width=0.9\linewidth]{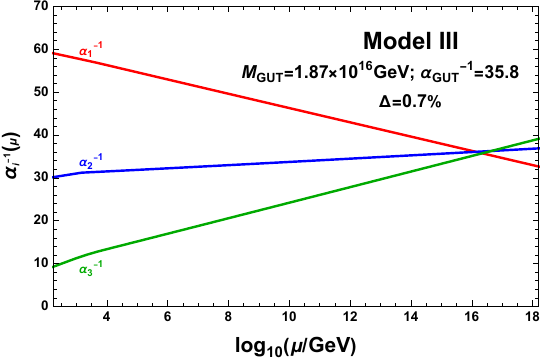}
    \caption{The evolution of two-loop gauge couplings in the Model III with vector-like particles $(XQ, \overline{XQ})$ and $(XD', \overline{XD'})$. The masses of the vector-like particles are set as follows: $M_{(XQ, \overline{XQ})}=1.5$ TeV for $(XQ, \overline{XQ})$, and $M_{(XD', \overline{XD'})}=5.5$ TeV for $(XD', \overline{XD'})$.}
    \label{fig:RGE_MIII}
\end{figure}

Second, for the renormalizable Yukawa couplings, we obtain that
the vector-like particles $(XQ, \overline{XQ})$ 
and $(XD', \overline{XD'})$ have masses
around $10^{11}$ GeV. In addition, the Lagrangian for the 
non-renormalizable Yukawa couplings is
\begin{eqnarray}
	-{\cal L} = \frac{1}{2M_{\rm Pl}} y_{XQ} XQ \overline{XQ} \Phi^2_1 
	+ \frac{1}{2M_{\rm Pl}} y_{XD'} XD' \overline{XD'} \Phi^2_2~,~
\end{eqnarray}
where $M_{\rm Pl}$ is the reduced Planck scale. Thus, 
we obtain that the vector-like particles $(XQ, \overline{XQ})$ 
and $(XD', \overline{XD'})$ have masses
around TeV scale. In this case, we can achieve 
the gauge coupling unification at $M_{\text{GUT}}=2.14\times 10^{16}$ GeV with a 0.7\% relative error. In Fig.~\ref{fig:RGE_MIII}, the vector-like particles are  introduced at $M_{(XQ, \overline{XQ})}=1.5$ TeV and $M_{(XD', \overline{XD'})}=5.5$ TeV. 
From the beta functions of these particles, $(XQ, \overline{XQ})$ modifies the RGE evolutionary trends of $\alpha_1^{-1}$, $\alpha_2^{-1}$ and $\alpha_3^{-1}$, while $(XD', \overline{XD'})$ alters the RGE trends of $\alpha_1^{-1}$ and $\alpha_3^{-1}$. 
The contribution of these two pairs of vector-like particles to $\alpha_1^{-1}$ is smaller than those to $\alpha_2^{-1}$ and $\alpha_3^{-1}$.  Moreover, the mass splitting between $(XQ, \overline{XQ})$ 
and $(XD', \overline{XD'})$ enables targeted mitigation of the relative error $\Delta$. As $\Delta$  quantifies the fractional separation of $\alpha_2$ and $\alpha_3$ from $\alpha_1$, the relative error decreases with the increasing mass of $(XD', \overline{XD'})$ and decreasing mass of  
$(XQ, \overline{XQ})$. For instance, when $(XD', \overline{XD'})$ 
has mass $10^5$ GeV and $(XQ, \overline{XQ})$ has mass 1.5 TeV, 
the error is reduced to 0.1\%. However, $(XQ, \overline{XQ})$ 
may not satisfy the LHC search constraints.

\section{Conclusions}

We constructed the generic high-quality axion models with anomalous $U(1)_X$ gauge symmetry 
and vector-like particles.
First, we briefly reviewed the gauge anomaly cancellations via the Green-Schwarz mechanism, discussed the $U(1)_X$ gauge symmetry breaking, as well as calculated the Nambu-Goldstone boson, PQ axion, and axion decay constant in  general. 
Because of the anomalous $U(1)_X$ gauge symmetry, we showed that 
the high-dimensional operators, which break the $U(1)_{PQ}$ global symmetry, 
can have dimension eleven or higher.
Therefore, the axion quality problem is solved. 
Interestingly, we only need to introduce two pairs of vector-like particles. 
To be concrete, we presented three specific models with two pairs of vector-like particles:
In Model I, these
vector-like particles belong to the fundamental and anti-fundamental representations of the $SU(5)$ model; in Model II, they belong to the anti-symmetric representation and
its Hermitian conjugate of the flipped $SU(5)$ model; in Model III, one pair of vector-like particles has the same quantum numbers as the quark doublet and its Hermitian conjugate, 
while the other pair has the same quantum numbers as the right-handed down-type quark and 
its Hermitian conjugate.
We showed that gauge anomalies in all three models can be canceled via the Green-Schwarz mechanism. To achieve gauge coupling unification, we need to introduce additional vector-like particles in Model I.
We found that gauge coupling unification is achieved at the unification scale around $10^{16}$ GeV with a relative error of less than 1\%. In particular, gauge coupling unification in Model II is achieved naturally with the smallest relative error of 0.1\%.

\section{Acknowledgments}
	
TL is supported in part by the National Key Research and Development Program of China Grant No. 2020YFC2201504, by the Projects No. 11875062, No. 11947302, No. 12047503, and No. 12275333 supported by the National Natural Science Foundation of China, by the Key Research Program of the Chinese Academy of Sciences, Grant No. XDPB15, by the Scientific Instrument Developing Project of the Chinese Academy of Sciences, Grant No. YJKYYQ20190049, and by the International Partnership Program of Chinese Academy of Sciences for Grand Challenges, Grant No. 112311KYSB20210012, by the 	HG and LW are supported in part by the Natural Science Basic Research Program of Shaanxi, Grant No. 2024JC-YBMS-039.
WZ is in part by the National Natural Science Foundation of China under grant no. 12405120, Start-up Funds for Young Talents of Hebei University (No.521100224226).

 \bibliography{Axion-LWZ}


\end{document}